\documentclass[
twocolumn,
amsmath,
amssymb,
aps,
prl,
]{revtex4-2}

\usepackage{bm}
\usepackage{mathrsfs}
\usepackage{hyperref}

\newcommand{\tens}[1]{\bm{\mathsf{#1}}}

\renewcommand{\i}{\ensuremath{\mathrm{i}}}
\renewcommand{\k}{\bm{k}}
\newcommand{\q}{\bm{q}}
\newcommand{\R}{\bm{R}}
\newcommand{\V}{\tens{V}}
\newcommand{\N}{\tens{N}}
\newcommand{\Nt}{\tilde{\N}}
\newcommand{\W}{\tens{\Omega}}
\renewcommand{\P}{\tens{Q}}
\newcommand{\Pt}{\tilde{\P}}
\newcommand{\Lin}{\mathcal{L}}
\newcommand{\Gr}{\mathcal{G}}

\begin{document}
\date{\today}
\title{greenWTE: Frequency-domain solver for the phonon Wigner Transport Equation with arbitrary heating using Green's functions}
\author{Laurenz Kremeyer}
\email{laurenz.kremeyer@mail.mcgill.ca}
\affiliation{Department of Physics, Centre for the Physics of Materials, McGill University, Montreal, Canada}
\author{Bradley J. Siwick}
\affiliation{Department of Physics, Centre for the Physics of Materials, McGill University, Montreal, Canada}
\affiliation{Department of Chemistry, McGill University, Montreal, Canada}
\author{Samuel Huberman}
\email{samuel.huberman@mcgill.ca}
\affiliation{Department of Chemical Engineering, McGill University, Montreal, Canada}
\affiliation{Department of Physics, McGill University, Montreal, Canada}
\begin{abstract}
Heat transport in solids is usually described in a phenomenological way via Fourier's law.
However, at small time and length scales in low-thermal-conductivity materials, a microscopic description is needed.
This is traditionally done by invoking the phonon Boltzmann Transport Equation (BTE)~\cite{ziman2001electrons}, which has been successful in describing transport in high-thermal-conductivity materials.
Recently, a generalization of the BTE, the Wigner Transport Equation, was developed to be able to microscopically describe transport in low-thermal-conductivity materials by taking into account not only phonon scattering processes but also phonon tunneling processes, so-called coherences~\cite{simoncelli2022wigner}.
We present a Python package that solves the Wigner Transport Equation (WTE) in spatial and temporal Fourier space for arbitrary source terms, thereby enabling the study of heat transport at small space and time scales.
This allows one to compute both static and dynamic thermal conductivities from bulk to nanoscale.
Beyond that, it can be used to study the response of phonon systems in materials to arbitrary heat sources~\cite{kremeyer2026transition}.
\end{abstract}

\maketitle

\section{Introduction}
There exists open-source code to determine the bulk thermal conductivity by solving the WTE for a static linear temperature gradient \cite{phono3py}.
Additionally, there exist closed-source implementations that solve the BTE in spatial and temporal Fourier space with arbitrary source terms using a Green's function approach~\cite{chiloyan2021green}.
Thus, we present a solver that fills this gap and supports microscopic thermal transport calculations in the WTE framework beyond the static limit.
Specifically, the solver is able to capture size and frequency effects, including nondiffusive effects, by allowing the user to choose arbitrary source terms.
By solving the WTE in spatial and temporal Fourier space, a bridge to common ultrafast and spatially patterned heating experiments is built.
In particular, the frequency-domain Green's function formulation naturally connects to ultrafast pump-probe experiments, where the heating is quasi-instantaneous and the subsequent thermal response is measured as a function of delay time~\cite{kremeyer2024ultrafast}.
Spatially periodic temperature gratings used in transient thermal grating experiments are directly modeled by our approach as well~\cite{ding2022observation, kremeyer2024ultrafast}.
The approach presented here is easily generalizable and can be applied to arbitrary experimental and device geometries.

Thermal conductivities are routinely calculated from first principles by solving the phonon BTE using open-source packages like Phono3py~\cite{phono3py}, ShengBTE~\cite{ShengBTE} or AlmaBTE~\cite{almaBTE}.
Complex crystals with large unit cells and low thermal conductivities need to be treated in the WTE framework, which is currently only implemented in Phono3py.
However, all three packages are limited to static bulk thermal conductivities.

\section{Current Functionality}
The formulation of the WTE in spatial and temporal Fourier space reads
\begin{equation}
\begin{split}
&-\i \omega \Nt({\k},{\q},\omega)+\i\Big[\W(\q),\Nt({\k},{\q},\omega)\Big] \\
&\quad+\frac{\i \k}{2}\Big\{\V(\q), \Nt({\k},{\q},\omega)\Big\} \\
&=\mathscr{F} \left\{\frac{\partial}{\partial t}\N({\R},{\q},t) \bigg|_{\mathrm{H}^{\mathrm{col}} }\right\} + \Pt({\k},{\q},\omega).
\end{split}
\end{equation}
Here $\omega$ and $\k$ are the temporal and spatial Fourier variables, $\q$ is the phonon wavevector, $\Nt$ is the Wigner distribution, $\W = \delta_{s,s'}\omega_{s}$ is the phonon frequency matrix, $\V$ is the velocity operator, $\mathscr{F}\left\{\frac{\partial}{\partial t}\N(\R,\q,t)\bigg|_{\mathrm{H}^{\mathrm{col}}}\right\}$ is the collision term and $\Pt(\k,\q,\omega)$ is the source term in Fourier space.
For a given pair of $(\k,\omega)$ we can rewrite this compactly as
\begin{equation}
\Lin(\k,\omega)\,\Nt(\k,\omega) = \Pt(\k,\omega).
\end{equation}
The inverse of the operator $\Lin(\k,\omega)$ is the Green's function $\Gr(\k,\omega)$, which allows one to solve for the Wigner distribution.
We currently implement two approaches to solve the presented system.
The first approach is a direct solver that fully inverts the operator $\Lin$ at each $(\k,\omega)$ point.
The determination of the Green's function is computationally expensive, but can subsequently be reused for the construction of the temperature response to user-chosen source terms in a computationally efficient manner.
The second approach solves the linear system $\Lin\,\Nt = \Pt$ directly for a user-determined source term using an iterative solver.
This avoids the expensive computation of the full Green's function, but requires one to solve the system from scratch for each source term.
The output of a greenWTE calculation is the Wigner distribution $\Nt(\k,\q,\omega)$ and derived integral quantities such as the heat flux and the thermal conductivity.

Possible source terms can be tailored to model different experimental heating conditions spatially like Gaussian or ring-like heating profiles~\cite{Varghese2023A,Jeong2021Transient}.
Further, energy can be selectively injected into specific regions of the Brillouin zone and different phonon modes.
Electron-phonon and phonon-phonon coupling calculations can be used to construct physically realistic source terms that are created by ultrafast laser radiation~\cite{Tong2021Toward}.

\section{Software Design}
We aim to provide a software package that is easy to use for researchers in the field of thermal transport and reuse the capabilities of Phono3py to compute all material-specific inputs to greenWTE.
In our theory, the majority of the computational workload is the direct or indirect inversion of the linear operator $\Lin$ for each point of the Brillouin zone.
greenWTE is a Python implementation that supports a NumPy~\cite{numpy} (CPU) and a CuPy~\cite{cupy} (GPU) backend to achieve high performance on modern hardware.
If a supported GPU is available, the linear-algebra-heavy parts of the code run on it.
While it is in principle possible to run the code on a CPU, a GPU is highly recommended to achieve reasonable performance in real-world scenarios.

\section{Research Impact Statement}
An outline of the research impact of greenWTE is provided in Ref.~\cite{kremeyer2026transition}, where we study the transition from population- to coherence-dominated nondiffusive thermal transport in low-thermal-conductivity materials.
Beyond that, the possibility to simulate pump-probe experiments by choosing source terms that model the nonequilibrium states created by ultrafast laser pulses will allow for a better interpretation of experimental results and a deeper understanding of microscopic thermal transport phenomena at short time and length scales.
A time-domain response function can be obtained by Fourier transforming the frequency-domain results and a direct comparison to time-domain experiments can be made.

\emph{Documentation}---Detailed documentation including installation instructions, tutorials, command-line interface options and API reference is available via \href{https://greenwte.readthedocs.io}{Read The Docs} on \url{https://greenwte.readthedocs.io}.

\emph{Acknowledgments}---L.~K. acknowledges support from a Fonds de Recherche du Québec-Nature et Technologies (FRQNT) Merit fellowship. B.~J.~S. acknowledges support from the NRC Quantum Sensors Challenge Program and the Canada Research Chairs program. S.~H. acknowledges support from the NSERC Discovery Grants Program under Grant No.~RGPIN-2021-02957 and FRQNT Nouveau Chercheur No.~341503.

\emph{AI Usage Disclosure}---Except for auto-completion features, the code in the repository is hand-written by the authors.
Generative AI tools were used to generate parts of the docstrings and to obtain suggestions for code improvements, bugfixes and variable naming consistency.
This manuscript was drafted without the use of generative AI tools; spell-checking, grammar-checking and wording suggestions from such tools were used.

\bibliography{main.maxbibnames3}
\end{document}